\documentclass[sigconf,nonacm]{acmart}

\usepackage[ruled,vlined,linesnumbered]{algorithm2e}
\usepackage{booktabs}
\usepackage{multirow}
\usepackage[normalem]{ulem}
\useunder{\uline}{\ul}{}
\usepackage[commandnameprefix=ifneeded,todonotes={textsize=tiny}]{changes}
\usepackage{enumitem}
\setlist{left=\parindent} 
\usepackage{bbold}
\usepackage[skins,listings]{tcolorbox}

\newtheorem{definition}{Definition}

\newcommand{\emphasis}[1]{\noindent\textbf{\textit{#1}}}

\newcommand{\sep}[1]{\footnotesize\textcolor{gray}{#1}}

\newcommand{\sys}{\textsc{UniBlocker}\xspace}

\usepackage{etoolbox}
\makeatletter
\patchcmd{\@setref}{\bfseries ??}{\bfseries \color{red} ??}{}{}
\patchcmd{\NAT@citex}{\bfseries ?}{\bfseries \color{red} ?}{}{}
\patchcmd{\NAT@citexnum}{\bfseries ?}{\bfseries \color{red} ?}{}{}
\ifdefined\HyRef@autosetref
  \patchcmd{\HyRef@autosetref}{\bfseries ??}{{\bfseries \color{red} ??}}{}{}
\fi
\makeatother

\begin{document}

\title{Towards Universal Dense Blocking for Entity Resolution}

\author{Tianshu Wang}
\orcid{0000-0002-1177-3901}
\affiliation{%
  \institution{Institute of Software, CAS}
  \institution{HIAS, UCAS}
  \country{China}
}
\email{tianshu2020@iscas.ac.cn}

\author{Hongyu Lin}
\affiliation{%
  \institution{Institute of Software, CAS}
  \country{China}
}
\email{hongyu@iscas.ac.cn}

\author{Xianpei Han}
\affiliation{%
  \institution{Institute of Software, CAS}
  \country{China}
}
\email{xianpei@iscas.ac.cn}

\author{Xiaoyang Chen}
\affiliation{%
  \institution{University of CAS}
  \country{China}
}
\email{chenxiaoyang19@mails.ucas.ac.cn}

\author{Boxi Cao}
\affiliation{%
  \institution{Institute of Software, CAS}
  \country{China}
}
\email{boxi2020@iscas.ac.cn}

\author{Le Sun}
\affiliation{%
  \institution{Institute of Software, CAS}
  \country{China}
}
\email{sunle@iscas.ac.cn}

\begin{abstract}
  Blocking is a critical step in entity resolution, and the emergence of neural network-based representation models has led to the development of dense blocking as a promising approach for exploring deep semantics in blocking. However, previous advanced self-supervised dense blocking approaches require domain-specific training on the target domain, which limits the benefits and rapid adaptation of these methods. To address this issue, we propose \sys, a dense blocker that is pre-trained on a domain-independent, easily-obtainable tabular corpus using self-supervised contrastive learning. By conducting domain-independent pre-training, \sys can be adapted to various downstream blocking scenarios without requiring domain-specific fine-tuning. To evaluate the universality of our entity blocker, we also construct a new benchmark covering a wide range of blocking tasks from multiple domains and scenarios. Our experiments show that the proposed \sys, without any domain-specific learning, significantly outperforms previous self- and unsupervised dense blocking methods and is comparable and complementary to the state-of-the-art sparse blocking methods.
\end{abstract}



\maketitle

\section{Introduction}
\label{sec:introduction}

\begin{figure}
  \centering
  \includegraphics[width=\columnwidth]{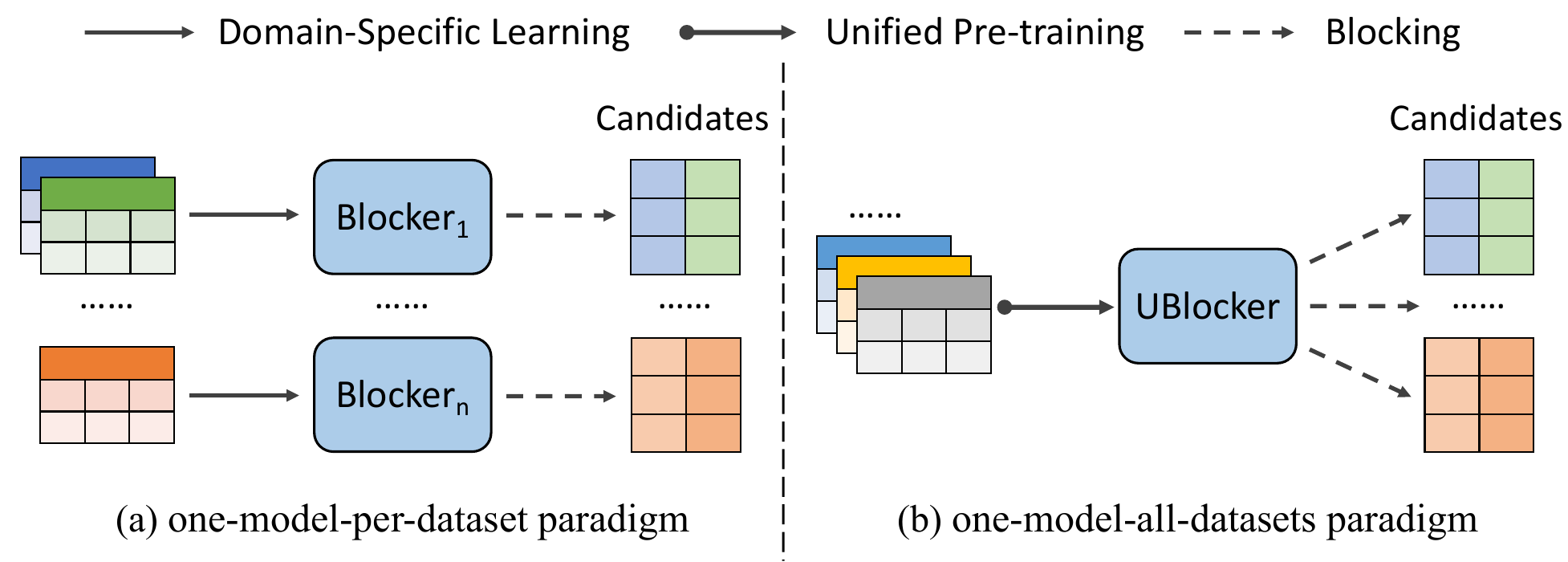}
  \caption{Illustration of \emph{one-model-per-dataset} paradigm and \emph{one-model-all-datasets} paradigm for dense blocking.}
  \label{fig:introduction}
\end{figure}

Entity resolution (ER)~\cite{fellegi-69-theor-recor-linkag,naumann2010introduction} aims to identify and merge duplicate records that refer to the same real-world entity, serving as a critical component in data cleaning, knowledge integration, and information management~\cite{ilyas-19-data-clean,christophides-21-overv-end-end-entit-resol-big-data,papadakis-21-four-gener-entit-resol}. A key challenge in entity resolution is the quadratic time complexity of pairwise record comparisons. To overcome this, the blocking-and-matching pipeline has become the mainstream solution in ER systems, where the blocking step groups records into small sets using computable features, and the matching step accurately matches records in each set with complex techniques. Blocking is essential for efficiently discarding incorrect options and reducing the time complexity in entity resolution.

The boost of neural network-based representation models has led to the development of \emph{dense blocking}, a promising approach to exploring deep semantics in the blocking step. Specifically, dense blocking utilizes an encoder, a neural network architecture, to convert entity records into dense semantic vectors. The distances between these vectors serve as a similarity measure between pairs of records. By performing nearest neighbor searches for each record, this technique generates a set of potential matching pairs or blocks. Compared to conventional syntactic-based sparse blocking approaches~\cite{papadakis-21-block-filter-techn-entit-resol}, dense blocking is more effective in modeling deep semantic information while maintaining scalability. As a result, dense blocking has become a research hotspot in the field of entity blocking~\cite{ebraheem-18-distr,zhang-20-autob,thirumuruganathan-21-deep,mugeni-22,wang-22-sudow,almagro-22-block-scl,zeakis-23-pre-train-embed-entit-resol}.

Unfortunately, recent studies~\cite{papadakis-22-how-entit-resol,zeakis-23-pre-train-embed-entit-resol} have shown that,
in contrast to supervised dense blocking methods~\cite{zhang-20-autob,brinkmann-23-sc-block},
self-supervised dense blocking methods~\cite{thirumuruganathan-21-deep,wang-22-sudow} do not offer substantial benefits and may be inferior to the direct use of pre-trained word or sentence embeddings. As illustrated in \autoref{fig:introduction}(a), we argue that this is because these methods learn \emph{domain-specific} models, which means that for each new blocking task, a unique representation model needs to be trained using the available records. Dense blocking methods in this paradigm demand considerable time and computational resources for each blocking.
Learning on domain-specific data can also lead to overfitting or collapse, making the outcomes not worth the effort.
However, the potential of self-supervised dense blocking learning from extensive data remains largely unexplored.

In this paper, we aim to extend the methodology of dense blocking from the one-model-per-dataset to a more generalized one-model-all-datasets paradigm, as shown in \autoref{fig:introduction}.
Motivated by recent advances in large language models~\cite{raffel2020exploring,brown2020language}, we develop a universal dense blocker, \sys, by exploiting the power of cross-domain self-supervised learning to pre-train on large amounts of tabular data. \autoref{fig:system} shows the overall pre-training procedure of \sys. \sys is built upon a Transformer-based architecture and trained with contrastive learning on GitTables~\cite{hulsebos-22-gittab}, a large-scale relational tabular corpus. Specifically, we filter and process the tables from this corpus into records, generate positive and negative pairs using data paraphrasing techniques at different levels, and obtain record representations by passing them through the encoder. The contrastive learning algorithm refines the model's capability to represent records.
Based on the aforementioned process, \sys, pre-trained on a heterogeneous, multi-domain, and multi-topic corpus of 1 million tables, acquires the capability to universally block entity collections across diverse domains and scenarios without further domain-specific learning.

To ascertain the effectiveness and universality of \sys, we gathered and constructed a new benchmark for universal blocking evaluation. The benchmark comprises 19 datasets spread across 9 domains and 4 scenarios, which enable a thorough evaluation of the performance of \sys and other blocking techniques.
Unlike previous studies that evaluate methods on partially labeled datasets and typically with identical schema records, this new benchmark emphasizes the ability of a blocking technique to transfer seamlessly across domains and scenarios without requiring extra resources.
As a result, the constructed benchmark evaluates entity blockers based on both effectiveness and scalability, considers both homogeneous and heterogeneous entity record collections~\cite{papadakis-21-four-gener-entit-resol}, and includes datasets of different scales from various domains.

Thorough experiments have been carried out on the constructed benchmark.
The experimental results highlight the effectiveness and universality of \sys, verifying the feasibility and benefits of universal dense blocking. Specifically, experiments show that \sys, without any additional domain-specific training, outperforms state-of-the-art (SOTA) self- and unsupervised dense blocking systems by almost 10\% in mean average precision. Furthermore, \sys is comparable to SOTA sparse blocking methods and is more effective in some complex and dirty scenarios where duplicate records are semantically similar. In addition, \sys demonstrated similar and even better efficiency when blocking large volumes of records.
Overall, these experimental results provide strong empirical support for the effectiveness, universality, and scalability of \sys.

\textbf{Contribution.} Our contributions can be summarized as follows:
\begin{itemize}
  \item We provide new insight into the limitations of existing self-supervised dense blocking methods that follow the one-model-per-dataset paradigm, and present our vision for one-model-all-datasets universal dense blocking.
  \item We propose a unified contrastive pre-training framework for developing the universal dense blocker, \sys. We design several optimizations to represent structural records better.
  \item We construct a new benchmark for universal blocking evaluation, which covers a wide range of blocking tasks from multiple domains and scenarios. This also lays the foundation for future research on automatic parameter search or good attribute identification for blocking~\cite{papadakis-22-how-entit-resol,paulsen-23-spark}.
  \item We conduct thorough experiments to demonstrate the effectiveness and universality of \sys by comparing it with the state-of-the-art dense and sparse blocking systems.
\end{itemize}


\section{Preliminaries and Background}
\label{sec:preliminaries}

In this section, we first introduce the problem definition and dense blocking. Then we discuss the limitations of blocking benchmarks.

\subsection{Problem Definition}

ER involves identifying records that refer to the same real-world entity, often called matches or duplicates. To avoid the quadratic number of comparisons with complicated entity matching algorithms, the blocking step is introduced in ER systems, which uses basic features to quickly scan all record pairs to achieve a reasonable trade-off between the number of comparisons and the loss of performance. Formally, the blocking step is defined in \autoref{pr:blocking}.

\begin{definition}[Blocking]
  \label{pr:blocking}
  Given entity record collections consisting of $n$ records $\mathcal{E} = \{e_i \mid 1 \le i \le n \}$, where each record $e$ is composed of attributes and values $e = \{(\text{attr}_i, \text{val}_i) \mid 1 \le i \le k\}$, the blocking step aims to generate candidate pairs $\mathcal{C} = \{(e_i, e_j) \} \subseteq \mathcal{E} \times \mathcal{E}$ to approximate the golden duplicate pairs $\mathcal{M} = \{(e_i, e_i^\prime)\} \subseteq \mathcal{E} \times \mathcal{E}$ under a given recall or comparison budget.
\end{definition}

Efficiency requirements for blocking require blocking algorithms to be scalable in terms of time and resource usage as records increase, and to generate a preferably linear quantity of candidates.

\subsection{Dense Blocking}

Dense blocking is an emerging technique for identifying potential matches in entity resolution with neural models. Unlike dense entity matching~\cite{mudgal-18-deep-learn-entit-match,barlaug-21-neural-networ-entit-match} where two records are simultaneously passed to a neural network, dense blocking transforms entity records into dense vectors independently and discovers potential matches through nearest neighbor search.
Compared to conventional sparse blocking methods, dense blocking allows for the modeling of deep semantic information, which reduces labor-intensive involvement~\cite{li-18-match,galhotra-21-beer,papadakis-22-how-entit-resol}. The technique has evolved from using context-free word embeddings~\cite{thirumuruganathan-18-reuse-adapt-entit-resol-trans-learn} to contextual pre-training language models~\cite{li-20-deep-entit-match-pre-train-languag-model,zeakis-23-pre-train-embed-entit-resol} and from supervised learning~\cite{zhang-20-autob} to self-supervised learning~\cite{thirumuruganathan-21-deep,wang-22-sudow}. The search for potential matches has also shifted from threshold-based~\cite{thirumuruganathan-18-reuse-adapt-entit-resol-trans-learn,zhang-20-autob} to cardinality-based~\cite{jain-21-deep,thirumuruganathan-21-deep}. Despite the significant progress made by self-supervised blocking methods~\cite{thirumuruganathan-21-deep,wang-22-sudow}, the one-model-per-dataset paradigm limits their application, making it necessary to develop the one-model-for-all-datasets paradigm.

\subsection{Benchmarks for Blocking}
\label{sec:benchmark limitations}

Unlike entity matching, the standard benchmarks for evaluating blocking methods are limited. Some studies~\cite{zhang-20-autob} have utilized non-public datasets, making experiment replication more challenging. Some other works~\cite{thirumuruganathan-21-deep} leverage entity matching datasets\cite{mudgal-18-deep-learn-entit-match} directly, which may introduce evaluation bias for blocking due to selection bias during dataset construction. The process of building an entity matching dataset without golden labels involves blocking and labeling. Sampling bias occurs when a particular blocking method is used to generate candidates, or when only a subset of candidates are labeled by an unspecified selection.
Despite numerous recent works on entity blocking~\cite{papadakis-22-how-entit-resol,paulsen-23-spark,zeakis-23-pre-train-embed-entit-resol}, the evaluation datasets used in these studies are limited to tables with identical schema. Therefore, existing benchmarks typically focus only on datasets with biases or in a specific scenario.
As a result, previous benchmarks cannot effectively evaluate the universal performance of entity blockers in a general scenario.


\section{\sys: A Universal Dense Blocker for Entity Blocking}
\label{sec:universal blocker}

In this section, we propose a contrastive pre-training framework to develop \sys, a universal dense blocker capable of blocking records across domains without domain-specific learning.

\subsection{Overview}
\begin{figure}
  \centering
  \includegraphics[width=\columnwidth]{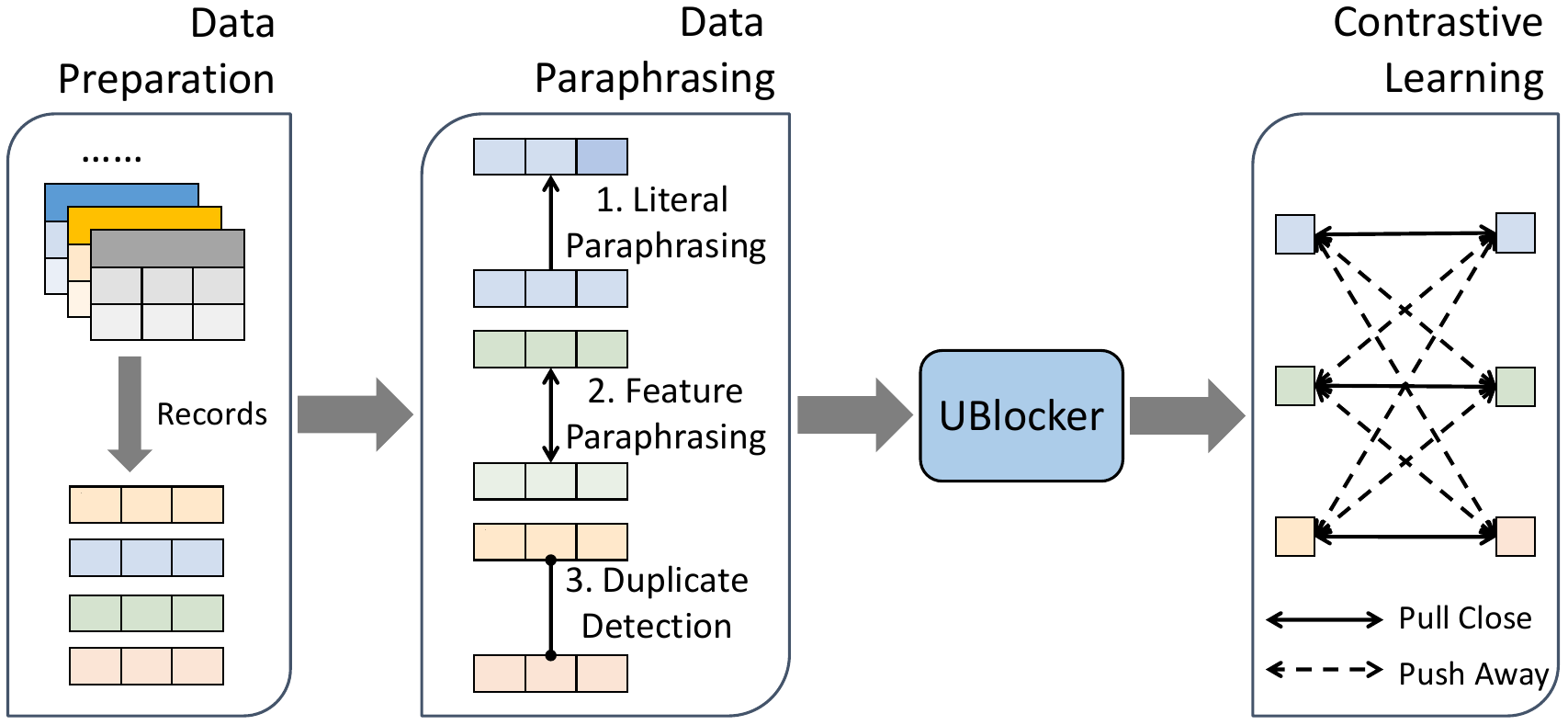}
  \caption{Overview of \sys pre-training procedure.}
  \label{fig:system}
\end{figure}

\autoref{fig:system} shows the overall pre-training procedure of \sys, which is a deep Transformer-based neural network trained on a large-scale relational tabular corpus using the contrastive learning algorithm.
Specifically, \sys starts by performing column detection and data cleaning to transform tables into sets of records. Subsequently, each record is transformed into a sequence of embeddings, and a Transformer-based encoder is employed to generate a dense semantic vector. To learn an effective record encoder, we leverage data paraphrasing techniques to generate positive and negative record pairs and contrastive learning is conducted by pulling positive pairs closer while pushing negative pairs further apart.

In the following, we will explain the process of acquiring the pre-training data from the tabular corpus. We will then describe the model architecture of \sys and demonstrate the approach by which \sys is trained using self-supervised contrastive learning on tabular records.

\subsection{Pre-training Data Conditioning}

The divergence between unstructured natural language and structured records (i.e., tuples of attributes and values) makes the direct application of pre-trained natural language models for table-related tasks suboptimal~\cite{dong-22-table-pre}. To fill this gap, in this paper, we propose to address this problem by pre-training \sys on large-scale relational tables to align the unstructured natural language models with structured and semi-structured records.

Specifically, we leverage GitTables~\cite{hulsebos-22-gittab} as our pre-training corpus because it is the first large table corpus derived from offline relational tables.
Despite the initial filtering conducted by the GitTables authors, we find that GitTables still contains a very large number of tables with various data types, including annotated data for machine learning, tabular data of statistics or features, log data, and so on. Since not all tables contain entities suitable for learning entity blocking, we used decision trees to learn a heuristic filtering algorithm based on column type detection to remove non-entity tables. Our filtering method heuristically determines the type of each column and then removes non-entity tables based on the results. More details about data conditioning, including column type detection and filtering, can be found in our supplementary materials.

After cleaning the corpus, we transform all tables into sets of entity records. Each row in a table is regarded as a record.
These sets of records are then fed into the \sys to learn a universal entity blocker, as illustrated below.

\begin{figure}
  \centering
  \includegraphics[width=\columnwidth]{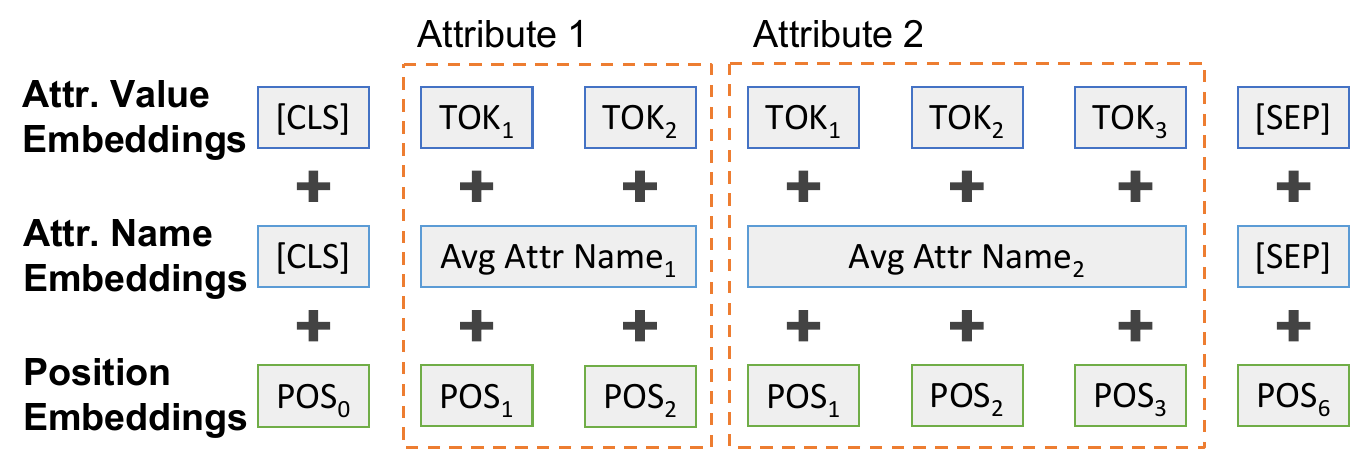}
  \caption{Record input representation. Record embeddings are the sum of token embeddings of attribute value, attribute name embeddings and relative position embeddings.}
  \label{fig:record embedding}
\end{figure}

\subsection{Encoder Architecture}

In this section, we present how we transform a record into its dense representation. Specifically, we propose a new record embedding technique that facilitates the conversion of structured records into a sequence of embeddings, allowing Transformer-based architectures to better model individual records for blocking. To achieve this, we transform the common tiled serialization into a combination of multiple embeddings. A Transformer-based encoder is then adapted to obtain the representation of the records.

\textbf{Record Embedding.} Serialization is a prerequisite for passing entity records to the Transformer architecture. In ER, it is common to serialize each record $e = {\{(\text{attr}_i, \text{val}_i)\}}_{1 \le i \le k}$ by concatenating attribute name and value tuples sequentially~\cite{li-20-deep-entit-match-pre-train-languag-model}. However, we have observed that this serialization technique is effective for matching but less effective for blocking. The reason for this is that blocking does not involve interactions between record pairs. Consequently, the aforementioned method can suffer from performance degradation when attribute values are missing or of limited length and the input sequence would mainly consist of attribute names. Additionally, the Transformer's use of position embeddings introduces the order of attributes as a significant factor. This has sparked debates about attribute shuffling in various studies~\cite{li-20-deep-entit-match-pre-train-languag-model,miao-21-rotom,wang-22-sudow}.

Based on these observations, we propose a record embedding method with bare-bone serialization, as shown in \autoref{fig:record embedding}. First, we concatenate all non-null attribute values of a record to form the input sequence: \( \text{val}_1\texttt{\char32}\text{val}_2\texttt{\char32}\ldots\texttt{\char32}\text{val}_k \), where $\texttt{\char32}$ stands for space.
We then incorporate attribute name information into token embeddings within a sequence by augmenting the embedding of each token with the average embedding of the corresponding attribute names.
Finally, to overcome the problem of misaligned focus on attributes, we propose a per-attribute position embedding method as a replacement for the original position embedding.
The per-attribute position embedding is obtained by resetting the position id at the beginning of each attribute value and converting it to embedding sequences.
The resulting serialized record embedding includes attribute name and relative position information.

\textbf{Transformer-based Encoder.} In recent years, Transformer-based pre-trained language models (PLMs) have attracted considerable attention and achieved state-of-the-art performance in numerous areas, including ER and DI\&P~\cite{li-20-deep-entit-match-pre-train-languag-model,jain-21-deep,wang-22-sudow}
, demonstrating the power of this architecture. Despite many practices in table pre-training~\cite{dong-22-table-pre}, none of them are trained for record representations and can be applied to blocking directly, so we employ the Transformer-based architecture and build on existing PLMs.

Given a record, we first transform it into its corresponding record embedding with the abovementioned technique. This embedding is then fed into a series of self-attention and feed-forward neural network modules.
Our goal in this study is to obtain dense record representations for blocking, so we leverage the encoder-only architecture. After the record undergoes transformation through the Transformer-based encoder, we use the average embeddings of all tokens in the last layer as the final record representation. For more information on the Transformer architecture, please refer to~\cite{vaswani2017attention}.

\begin{table}
  \centering
  \caption{Literal Data Paraphrasing Actions.}\label{tab:data paraphrasing}
  \resizebox{\columnwidth}{!}{%
    \begin{tabular}{ccc}
      \toprule
      Level & Action                                                                      & Description                                                                                \\ \midrule
      Char. & substitute                                                                  & Simulate keyboard distance and OCR engine error                                            \\
            & \begin{tabular}[c]{@{}c@{}}insert, substitute, \\ swap, delete\end{tabular} & Apply paraphrasing randomly                                                                \\ \midrule
      Token & substitute                                                                  & Find suitable word from dictionary                                                         \\
            & \begin{tabular}[c]{@{}c@{}}insert, split, crop, \\ swap, delete\end{tabular}& Apply paraphrasing randomly                                                                \\ \midrule
      Cell  & substitute                                                                  & Substitute attribute name from dictionary                                                  \\
            & crop, delete                                                                & Apply paraphrasing randomly                                                                \\ \bottomrule
    \end{tabular}%
  }
\end{table}

\subsection{Data Paraphrasing for Positive Pair Generation}
One of the main challenges in self-supervised learning is to generate positive samples that are optimal for downstream tasks~\cite{xiao2021should}. To address this challenge, we leverage two data paraphrasing techniques, namely literal paraphrasing and feature paraphrasing, from different perspectives to generate positive pairs. We also introduce duplicate detection that identifies potential positives in the original data, preventing performance degradation due to false negatives.

\textbf{Literal Paraphrasing} The presence of redundant, missing, misplaced, misspelled, and noisy records in entity resolution has received widespread attention~\cite{nie-19-deep-sequen-sequen-entit-match,fu-20-hierar-match-networ-heter-entit-resol}. Taking advantage of the record embedding design, we can focus on simulating the dirtiness of records in the real scenario. Based on this intuition, we perform literal paraphrasing at three levels: character, token, and attribute. The paraphrasing actions are summarized in \autoref{tab:data paraphrasing}.

\textbf{Feature Paraphrasing.}
Our approach differs from Sudowoodo that applies the cutoff operation at the embedding level~\cite{wang-22-sudow}, while we employ dropout for feature paraphrasing across all Transformer layers. Dropout is a technique that prevents overfitting of neural networks during training by randomly deactivating neurons with a probability $p$. When deployed, the encoder creates a slightly varied feature representation of the same input.
Therefore, we encode identical records with dropout twice to obtain different representations serving as positive samples for training, which has been confirmed to be highly effective by prior research~\cite{gao2021simcse}.

\textbf{Duplicate Detection.}
In addition to manual data paraphrasing, existing ``duplicate'' records can also serve as positive pairs for training. Two records are considered ``duplicate'' if they share high similarity and can be transformed into each other by a few steps of literal paraphrasing. To avoid performance degradation due to false negatives, we propose a heuristic method to detect duplicates. This method involves computing the similarities between records using various string similarity measures~\cite{doan2012principles}. For each pair of records in the batch, if any of the computed similarities exceed a predetermined threshold $s$, we consider them to be positive pairs.

From another perspective, duplicate detection can be seen as weak supervised learning based on sparse similarity measures.
Since the blocking phase cannot afford such computationally intensive measures, models can be trained in this way to generate similar representations for similar records across different measures.

\subsection{Model Training}

In the following, we will describe how we train \sys with the aforementioned techniques. Formally, given a batch of $N$ records donated as $E = {\{e_i\}}_{1 \le i \le N}$, we generate positive samples for each of them by applying literal paraphrasing, resulting in $E^\prime = {\{e_i^\prime\}}_{1 \le i \le N}$. We then apply record embedding to these $2N$ instances and feed them into the encoder with feature paraphrasing, obtaining their representations $\boldsymbol{Z} = {\{\boldsymbol{z}_i\}}_{1 \le i \le N}$ and $\boldsymbol{Z}^\prime = {\{\boldsymbol{z}_i^\prime\}}_{1 \le i \le N}$. For the constructed pairs from these $2N$ records, we consider the pairs generated through data paraphrasing and duplicate detection as positive pairs, while the remaining pairs in the batch are regarded as negative pairs. Finally, since each instance may have multiple positive samples, we adapt Circle Loss~\cite{sun2020circle} as our loss function. Suppose for each sample $e$ and its representation $\boldsymbol{z}$ we have $J$ positives and $K$ negatives, where $J + K = N$. We denote the similarity of their paraphrased version representations to $\boldsymbol{z}$ are ${\{s_p^j\}}_{1 \le j \le J}$ and ${\{s_n^k\}}_{1 \le k \le K}$ respectively. The loss function is described as:

\begin{equation}
  \label{eq:CircleLoss}
  \mathcal{L}_{\text {Circle}} = \sum_{i=1}^{N}\log \left[1+\sum_{j=1}^{J} \sum_{k=1}^{K} \exp \left(\gamma\left(s_{n}^{k}-s_{p}^{j}+m\right)\right)\right]
\end{equation}
where $\gamma$ is a scaling factor acting similar to $\tau$ in InfoNCE loss~\cite{chen2020simple} and $m$ is a margin for better similarity separation.


\section{Benchmarking Universal Blocking}
\label{sec:benchmark}

This section presents a new benchmark and evaluation to overcome the limitations of blocking benchmarks, as discussed in \autoref{sec:benchmark limitations}. This aims to assess the universality of blocking methods.

\begin{table}
  \caption{Datasets of our proposed benchmark. We superscript datasets appeared in previous literature with asterisk.}
  \label{tab:benchmark}
  \resizebox{\linewidth}{!}{
    \begin{tabular}{ccccc}
      \toprule
      Dataset                                  & Domain      & Size            & \# Attr.   & \# Pos   \\ \midrule \multicolumn{5}{c}{Effectiveness} \\ \midrule
      census                                   & census      & 841             & 5         & 344       \\
      cora                                     & citation    & 1879            & 18        & 64578     \\
      notebook                                 & notebook    & 343             & 13        & 2152      \\
      notebook2                                & notebook    & 1661            & 1         & 2814      \\
      altosight                                & product     & 835             & 4         & 4082      \\
      altosight2                               & product     & 2006            & 5         & 4393      \\ \cmidrule(lr){1-5}
      fodors-zagats\_homo\textsuperscript{*}   & restaurant  & 553, 331        & 6         & 112       \\
      dblp-acm\textsuperscript{*}              & citation    & 2616, 2294      & 4         & 2224      \\
      dblp-scholar\textsuperscript{*}          & citation    & 2616, 64263     & 4         & 5438      \\
      abt-buy\_homo\textsuperscript{*}         & product     & 1081, 1092      & 2         & 1098      \\
      walmart-amazon\_homo\textsuperscript{*}  & electronics & 2554, 22074     & 5         & 1154      \\ \cmidrule(lr){1-5}
      fodors-zagats\_heter                     & restaurant  & 553, 331        & 4,5       & 112       \\
      imdb-dbpedia\textsuperscript{*}          & movies      & 27615, 23182    & 4,7       & 22863     \\
      amazon-google\textsuperscript{*}         & product     & 1363, 3226      & 4,4       & 1300      \\
      abt-buy\_heter                           & product     & 1081, 1092      & 2,3       & 1098      \\
      walmart-amazon\_heter                    & electronics & 2554, 22074     & 16,21     & 1154      \\ \cmidrule(lr){1-5}
      movies                                   & movies      & 18984           & 14,31,10  & 4135      \\ \midrule \multicolumn{5}{c}{Scalability} \\ \midrule
      songs                                    & musics      & 1E+6            & 7         & -         \\
      citeseer-dblp                            & citation    & 1.8E+6, 2.5E+6  & 6         & -         \\
      \bottomrule
    \end{tabular}
  }
\end{table}


\subsection{Overview of Proposed Benchmark}

We propose a new benchmark that prioritizes diversity while ensuring accuracy by collecting public datasets with full golden labels. Specifically, we focus on diversity across multiple factors:
\begin{itemize}
\item \textbf{Diverse Scenarios.} The entity resolution scenario includes two aspects, the number of collections and schema heterogeneity. Unlike previous benchmarks, we cover the case where the input is from a single or double collection of records. For input from a single collection, which is often known as dirty ER or deduplication, there is no schema heterogeneity and all records have the same attributes. For input from two collections (i.e., record linkage), we consider two scenarios where the schema is homogeneous or heterogeneous.
Although schema-agnostic methods can handle both scenarios, little work has been done on the performance differences between these two scenarios. Additionally, we cover the scenario where entity collections come from multiple sources, which is more challenging due to schema heterogeneity.
\item \textbf{Diverse Domains.} For each scenario, we aim to include datasets from as many different domains as possible. The inconsistency of duplicate records from different domains places different demands on blocking methods.
\item \textbf{Diverse Properties.} In addition to the domains, the properties of datasets are also critical factors that can affect the performance of different methods. For instance, a table with a single attribute may be considered textual, whereas a table with many attributes may pose a challenge for finding key attributes. Additionally, the similarity between duplicate and non-duplicate records and the average number of duplicates per record are internal properties of the dataset that may influence method performance.
\end{itemize}

By incorporating the diversity of these factors, our benchmark provides a comprehensive evaluation of blocking methods.
This also lays the foundation for future research on automatic parameter search or good attribute identification for blocking~\cite{papadakis-22-how-entit-resol,paulsen-23-spark}.

\autoref{tab:benchmark} provides an overview of the datasets included in our benchmark, which comprises a total of 17 datasets from 9 different domains. These datasets include 6 from one source, 10 from two sources and 1 from three sources. For the datasets sourced from two sources, we include 5 homogeneous and 5 heterogeneous datasets. The number of attributes per entity record ranges from 1 to 31, with an average number of duplicates per record ranging from 0.02 to 1. Different ``notebook'' and ``altosight'' datasets originate from different years of the SIGMOD Programming Contest. After preprocessing and labeling, most datasets are found to be homogeneous, especially those from DeepMatcher~\cite{mudgal-18-deep-learn-entit-match}. For the heterogeneous datasets, we include both artificially heterogeneous datasets, constructed by renaming and merging attributes as \cite{wang-21-macham}, and naturally heterogeneous datasets where raw data was available. For example, we constructed ``fodors-zagats\_heter'' and ``abt-buy\_heter'' by renaming and merging attributes. In contrast, ``walmart-amazon\_heter'' comprises the original heterogeneous records with more than 15 attributes and labels from ``walmart-amazon\_homo''. The multi-source ``movies'' dataset is a merge of three datasets (IMDB-TMDB, IMDB-TVDB, and TMDB-TVDB) obtained from JedAIToolkit~\cite{papadakis-20-three-entit-resol-jedai}. Details of the datasets and processing can be found in the supplementary material.

\subsection{Evaluation}
\label{sec:evaluation}

The evaluation metrics commonly used to assess the effectiveness of entity blocking methods, which are also used in our benchmark, are pair completeness (PC) and pair quality (PQ). These metrics are widely employed in the literature and provide a means to quantify the performance of entity blocking methods:

\begin{enumerate}
  \item Pair completeness (also known as recall), is the fraction of true matched pairs identified: $\mathrm{PC} = {\lvert \mathcal{C} \cap \mathcal{M} \rvert}/{\lvert \mathcal{M} \rvert}$
  \item Pair quality (also called precision), is the fraction of matched pairs among the candidate pairs: $\mathrm{PQ} = {\lvert \mathcal{C} \cap \mathcal{M} \rvert}/{\lvert \mathcal{C} \rvert}$
\end{enumerate}

Balancing pair completeness and pair quality is a challenging trade-off, as it is difficult to achieve both simultaneously under efficiency constraints. Since the matching phase provides further assess to pair quality and discarded pairs cannot be recovered, the blocking phase typically prioritizes pair completeness. However, the variability in the number of candidate pairs generated by different blocking methods and the complexity of the internal parameters make it difficult to fairly compare the performance of different methods. To address this issue, a configuration optimization evaluation setting~\cite{papadakis-22-how-entit-resol} has been proposed, which aims to tune all parameters of a given method to achieve a threshold for pair completeness. In this paper, we also follow this setting to set the threshold of pair completeness to 90\%. In practice, however, it is unrealistic to search for all parameters in a limited amount of time due to the missing golden labels and the huge parameter space. Therefore, we stick default parameters for all methods to assess their universality across all datasets. In addition, for cardinality-based methods using k-nearest neighbor (kNN) search, we establish an upper bound on k of 100 as a comparison budget and check the candidates in nearest neighbor order. We evaluate the blockers based on whether they can achieve pair completeness greater than 90\%. If both blockers meet this criterion, we compare their maximum pair quality. Otherwise, we compare their pair completeness. This approach allows us to assess the effectiveness of the blockers under different conditions and provide a comprehensive evaluation of their performance.

In addition, we leverage mean average precision (mAP) to evaluate the overall performance of kNN approaches. mAP, which is the area under the precision-recall curve (AUC-PR), is calculated as $\sum_{i} \left(\mathrm{R}_i - \mathrm{R}_{i-1}\right)\mathrm{P}_i$ where $\mathrm{R}_i$ and $\mathrm{P_i}$ represent PC and PQ at $k = i$.

\section{Experiments}
\label{sec:experiments}

\begin{table*}[]
  \caption{Overall results of different dense blocking methods. As discussed in \autoref{sec:evaluation}, if both methods exceed the preset PC threshold (90\%), we evaluate their PQ, otherwise, we compare their PC. K is the number of nearest neighbors needed to reach the reported PC and \textcolor{gray}{$\Delta\text{mAP}$} is the change after domain-specific fine-tuning. The best results are highlighted in bold.}
  \label{tab:dense}
  \centering
  \resizebox{0.85\textwidth}{!}{%
    \begin{tabular}{@{}rcccccccccccccccc@{}}
      \toprule
      \multicolumn{1}{l}{}         & \multicolumn{8}{c}{Domain-Specific Learning}                                                                                   & \multicolumn{8}{c}{Unified Pre-training}                                                                                                         \\ \cmidrule(lr){2-9} \cmidrule(lr){10-17}
      \multicolumn{1}{l}{}         & \multicolumn{4}{c}{DeepBlocker}                                       & \multicolumn{4}{c}{Sudowoodo}                          & \multicolumn{4}{c}{STransformer}                                  & \multicolumn{4}{c}{\sys}                                        \\ \cmidrule(lr){2-5} \cmidrule(lr){6-9} \cmidrule(lr){10-13} \cmidrule(lr){14-17}
      \multicolumn{1}{c}{Datasets} & mAP             & \multicolumn{2}{c}{PC / PQ}                 & K     & mAP      & \multicolumn{2}{c}{PC / PQ}       & K       & mAP               & \multicolumn{2}{c}{PC / PQ}                & K       & mAP \sep{$\Delta\text{mAP}$}             & \multicolumn{2}{c}{PC / PQ}                 & K     \\ \cmidrule(lr){1-1} \cmidrule(lr){2-5} \cmidrule(lr){6-9} \cmidrule(lr){10-13} \cmidrule(lr){14-17}
      census                       & 9.83            & \multicolumn{2}{c}{90.41 / 1.44}            & 33    & 8.73     & \multicolumn{2}{c}{90.12 / 1.43}  & 39      & 12.32             & \multicolumn{2}{c}{91.86 / 7.03}           & 9       & \textbf{16.04} \sep{-0.05}  & \multicolumn{2}{c}{\textbf{90.99 / 15.14}}  & 5     \\
      cora                         & 41.84           & \multicolumn{2}{c}{66.49 / 29.48}           & 100   & 27.29    & \multicolumn{2}{c}{46.47 / 25.58} & 100     & 50.80             & \multicolumn{2}{c}{\textbf{77.64 / 37.54}} & 100     & \textbf{51.21} \sep{+0.03} & \multicolumn{2}{c}{77.19 / 37.39}           & 100   \\
      notebook                     & 27.20           & \multicolumn{2}{c}{90.29 / 7.67}            & 97    & 29.69    & \multicolumn{2}{c}{90.20 / 10.22} & 84      & 28.49             & \multicolumn{2}{c}{90.01 / 10.50}          & 84      & \textbf{43.36} \sep{+0.83} & \multicolumn{2}{c}{\textbf{90.38 / 16.90}}  & 52    \\
      notebook2                    & 40.54           & \multicolumn{2}{c}{90.05 / 2.69}            & 62    & 33.48    & \multicolumn{2}{c}{86.57 / 2.25}  & 100     & 34.39             & \multicolumn{2}{c}{90.12 / 2.28}           & 86      & \textbf{41.55} \sep{+0.01} & \multicolumn{2}{c}{\textbf{90.37 / 3.97}}   & 52    \\
      altosight                    & 8.94            & \multicolumn{2}{c}{52.96 / 3.13}            & 100   & 24.77    & \multicolumn{2}{c}{85.82 / 6.33}  & 100     & 26.12             & \multicolumn{2}{c}{90.00 / 7.72}           & 85      & \textbf{38.83} \sep{-0.20} & \multicolumn{2}{c}{\textbf{90.32 / 13.74}}  & 51    \\
      altosight2                   & 12.56           & \multicolumn{2}{c}{46.93 / 1.18}            & 100   & 16.59    & \multicolumn{2}{c}{53.83 / 1.86}  & 100     & 22.99             & \multicolumn{2}{c}{81.72 / 2.44}           & 100     & \textbf{29.04} \sep{-0.02} & \multicolumn{2}{c}{\textbf{89.80 / 2.81}}   & 100   \\ \cmidrule(lr){1-1} \cmidrule(lr){2-5} \cmidrule(lr){6-9} \cmidrule(lr){10-13} \cmidrule(lr){14-17}
      fodors-zagats\_homo          & \textbf{60.51}  & \multicolumn{2}{c}{\textbf{100.00 / 21.01}} & 1     & 59.90    & \multicolumn{2}{c}{99.11 / 20.83} & 1       & 53.42             & \multicolumn{2}{c}{93.75 / 9.85}           & 2       & \textbf{60.51} \sep{+0.00} & \multicolumn{2}{c}{\textbf{100.00 / 21.01}} & 1     \\
      dblp-acm                     & 90.29           & \multicolumn{2}{c}{97.39 / 82.80}           & 1     & 90.02    & \multicolumn{2}{c}{97.21 / 82.65} & 1       & 88.21             & \multicolumn{2}{c}{95.28 / 81.00}          & 1       & \textbf{91.88} \sep{-0.06} & \multicolumn{2}{c}{\textbf{99.19 / 84.33}}  & 1     \\
      dblp-scholar                 & 67.85           & \multicolumn{2}{c}{90.35 / 20.52}           & 9     & 57.40    & \multicolumn{2}{c}{90.01 / 2.16}  & 85      & 69.85             & \multicolumn{2}{c}{91.49 / 23.38}          & 8       & \textbf{73.12} \sep{+0.31} & \multicolumn{2}{c}{\textbf{91.55 / 31.19}}  & 6     \\
      abt-buy\_homo                & 43.55           & \multicolumn{2}{c}{90.06 / 1.02}            & 90    & 69.79    & \multicolumn{2}{c}{90.52 / 15.31} & 6       & 42.83             & \multicolumn{2}{c}{90.52 / 7.07}           & 13      & \textbf{79.80} \sep{+0.36} & \multicolumn{2}{c}{\textbf{93.16 / 31.51}}  & 3     \\
      walmart-amazon\_homo         & 39.37           & \multicolumn{2}{c}{90.64 / 2.73}            & 15    & 44.13    & \multicolumn{2}{c}{90.29 / 4.08}  & 10      & 30.73             & \multicolumn{2}{c}{90.03 / 1.51}           & 27      & \textbf{53.22} \sep{-0.44} & \multicolumn{2}{c}{\textbf{93.33 / 14.06}}  & 3     \\ \cmidrule(lr){1-1} \cmidrule(lr){2-5} \cmidrule(lr){6-9} \cmidrule(lr){10-13} \cmidrule(lr){14-17}
      fodors-zagats\_heter         & \textbf{60.51}  & \multicolumn{2}{c}{\textbf{100.00 / 21.01}} & 1     & 56.07    & \multicolumn{2}{c}{92.86 / 19.51} & 1       & 20.69             & \multicolumn{2}{c}{90.18 / 0.29}           & 65      & 59.35          \sep{+0.00} & \multicolumn{2}{c}{98.21 / 20.64}           & 1     \\
      imdb-dbpedia                 & 0.00            & \multicolumn{2}{c}{0.00 / 0.00}             & 100   & 8.42     & \multicolumn{2}{c}{43.63 / 0.36}  & 100     & 8.07              & \multicolumn{2}{c}{46.06 / 0.38}           & 100     & \textbf{23.93} \sep{+2.19} & \multicolumn{2}{c}{\textbf{59.53 / 0.49}}   & 100   \\
      amazon-google                & 43.98           & \multicolumn{2}{c}{90.15 / 1.12}            & 77    & 52.29    & \multicolumn{2}{c}{90.23 / 7.17}  & 12      & 46.20             & \multicolumn{2}{c}{90.46 / 6.64}           & 13      & \textbf{62.31} \sep{-0.17} & \multicolumn{2}{c}{\textbf{90.38 / 17.24}}  & 5     \\
      abt-buy\_heter               & 45.18           & \multicolumn{2}{c}{90.06 / 1.43}            & 64    & 71.73    & \multicolumn{2}{c}{90.79 / 18.43} & 5       & 47.17             & \multicolumn{2}{c}{91.70 / 9.31}           & 10      & \textbf{80.89} \sep{-0.01} & \multicolumn{2}{c}{\textbf{93.16 / 31.51}}  & 3     \\
      walmart-amazon\_heter        & 25.88           & \multicolumn{2}{c}{84.84 / 0.38}            & 100   & 28.57    & \multicolumn{2}{c}{90.12 / 0.67}  & 61      & \textbf{32.47}    & \multicolumn{2}{c}{\textbf{90.12 / 1.94}}  & 21      & 30.18          \sep{-0.65} & \multicolumn{2}{c}{90.03 / 0.75}            & 54    \\ \cmidrule(lr){1-1} \cmidrule(lr){2-5} \cmidrule(lr){6-9} \cmidrule(lr){10-13} \cmidrule(lr){14-17}
      movies                       & 7.53            & \multicolumn{2}{c}{89.84 / 0.25}            & 100   & 2.62     & \multicolumn{2}{c}{76.71 / 0.26}  & 100     & 3.06              & \multicolumn{2}{c}{68.54 / 0.19}           & 100     & \textbf{8.07}  \sep{+0.02} & \multicolumn{2}{c}{\textbf{90.04 / 2.20}}   & 13    \\ \cmidrule(lr){1-1} \cmidrule(lr){2-5} \cmidrule(lr){6-9} \cmidrule(lr){10-13} \cmidrule(lr){14-17}
      Mean                         & 36.80           & \multicolumn{2}{c}{80.03 / 11.64}           & 61.76 & 40.09    & \multicolumn{2}{c}{82.62 / 12.89} & 53.24   & 36.34             & \multicolumn{2}{c}{85.85 / 12.30}          & 48.47   & \textbf{49.60} \sep{+0.13} & \multicolumn{2}{c}{\textbf{89.86 / 20.29}}  & 32.35 \\ \bottomrule
    \end{tabular}%
  }
\end{table*}


In the following section, we conduct thorough experiments to determine the effectiveness and universality of \sys, as well as to identify its strengths and weaknesses in comparison to sparse blocking methods.
We first evaluate the performance of \sys against both domain-specific and unified sentence-based dense blocking methods. Then we compare \sys with sparse blocking methods. Finally, we conduct ablation studies.

\subsection{Experimental Setup}

The baselines involved in our experiments include:

\begin{itemize}
\item \textbf{DeepBlocker}~\cite{thirumuruganathan-21-deep}: DeepBlocker is a design space exploration for self-supervised blocking solutions. We use the best overall solution Autoencoder for comparison.
\item \textbf{Sudowoodo}~\cite{wang-22-sudow}: Sudowoodo is a contrastive, PLMs-based, self-supervised learning framework used to solve various data integration and preparation tasks, including blocking.
\item \textbf{STransformer}~\cite{reimers2019sentence}: STransformer is a sentence embedding framework trained on sentence pairs using siamese networks. It's the best blocking solution with pre-trained embeddings~\cite{zeakis-23-pre-train-embed-entit-resol}.
\item \textbf{Blocking Workflows}~\cite{papadakis-20-three-entit-resol-jedai}: Blocking Workflows is the general term for the sparse blocking pipeline, including block building, block cleaning, and comparison cleaning.
\item \textbf{Sparkly}~\cite{paulsen-23-spark}: Sparkly is a kNN blocker that uses Lucene to perform top-k blocking. The well-known tf/idf measure has shown impressively superior performance in \cite{paulsen-23-spark}.
\end{itemize}

We implemented \sys using Python 3.10, Pytorch 1.13, and the Transformers~\cite{wolf-etal-2020-transformers} library. We chose to use the all-mpnet-base-v2 from STransformer~\cite{reimers2019sentence} as the initial pre-training weight and set the maximum sequence length to 256. We set the probability of applying literal paraphrasing actions to 0.01, the dropout probability for feature paraphrasing to 0.15, and the similarity threshold $s$ for duplicate detection to 0.85. The scaling factor $\gamma$ and margin $m$ of the Circle Loss are the default 80 and 0.4, respectively. We used the AdamW optimizer with a learning rate of 1e-5, a linear learning rate schedule without warm-up steps, and a batch size of 128 for pre-training. The total number of pre-training steps is 10,000.

For all baselines, we followed their proposed implementations and settings.
As discussed in \autoref{sec:evaluation}, we kept consistent parameters for each method across all datasets to assess universality. For Blocking Workflows, we adhered to the DBW setting~\cite{papadakis-22-how-entit-resol}. For Sparkly, we employed the default Okapi BM25 scoring function and the 3-gram tokenizer. In terms of dense blocking baselines, we only serialized record values for effectiveness, using FAISS as the indexer.
If the dataset contains more than two entity collections, we merge them into a single one to convert it to dirty ER blocking.

\subsection{Comparison with Dense Blocking Methods}

\subsubsection*{\sys vs Domain-Specific Methods.}
To assess the effectiveness and universality of \sys, we first compare \sys with SOTA domain-specific self-supervised dense blocking methods.
Experimental results presented in \autoref{tab:dense} indicate that \sys outperforms previous domain-specific dense blocking methods by almost 10\% mAP on average.
In contrast to the lack of a clear winner between DeepBlocker and Sudowoodo, \sys consistently delivers the best performance among them on almost all datasets, further proving its effectiveness by reducing the number of nearest neighbors required.
These results indicate that \sys not only addresses the problem of limited downstream data, but also eliminates the need for domain-specific model training. Pre-training allows deep models to reach their full potential and learn a more general representation of the structured records.
In conclusion, our experimental results provide compelling evidence for the advantage of self-supervised dense blocking learning on large-scale open-domain data.

\emphasis{Finding 1. \sys outperforms SOTA domain-specific dense blockers, showing the feasibility of developing universal dense blockers via a unified domain-independent pre-training.}

\begin{table*}[]
  \caption{Overall results of sparse, dense blocking methods and their ensemble. The best results are emphasized in bold and the better ensemble results are underlined. Backslash (\textbackslash) indicates runtime errors.}
  \label{tab:dense-vs-sparse}
  \resizebox{0.80\textwidth}{!}{%
    \begin{tabular}{rcccccccccccccc}
      \toprule
      \multicolumn{1}{l}{}  & \multicolumn{6}{c}{Sparse}                                                                                        & \multicolumn{4}{c}{Dense}                                             & \multicolumn{4}{c}{Ensemble}                                         \\ \cmidrule(lr){2-7} \cmidrule(lr){8-11} \cmidrule(lr){12-15}
      \multicolumn{1}{l}{}  & \multicolumn{2}{c}{Blocking Workflows}     & \multicolumn{4}{c}{Sparkly\textsubscript{3}}                         & \multicolumn{4}{c}{\sys}                                        & \multicolumn{4}{c}{\sys + Sparkly\textsubscript{3}}                             \\ \cmidrule(lr){2-3} \cmidrule(lr){4-7} \cmidrule(lr){8-11} \cmidrule(lr){12-15}
      Datasets              & \multicolumn{2}{c}{PC / PQ}                & mAP            & \multicolumn{2}{c}{PC / PQ}                 & K     & mAP             & \multicolumn{2}{c}{PC / PQ}                 & K     & mAP             & \multicolumn{2}{c}{PC / PQ}             & K        \\ \cmidrule(lr){1-1} \cmidrule(lr){2-3} \cmidrule(lr){4-7} \cmidrule(lr){8-11} \cmidrule(lr){12-15}
      census                & \multicolumn{2}{c}{45.93 / 11.10}          & \textbf{25.55} & \multicolumn{2}{c}{\textbf{95.06 / 21.95}}  & 4     & 16.04           & \multicolumn{2}{c}{90.99 / 15.14}           & 5     & 18.11           & \multicolumn{2}{c}{95.93 / 17.66}       & 4        \\
      cora                  & \multicolumn{2}{c}{\textbf{97.51 / 65.32}} & 49.60          & \multicolumn{2}{c}{74.06 / 37.94}           & 100   & \textbf{51.21}  & \multicolumn{2}{c}{\textbf{77.19 / 37.39}}  & 100   & \underline{51.79}     & \multicolumn{2}{c}{\underline{82.39 / 28.91}} & 100      \\
      notebook              & \multicolumn{2}{c}{87.45 / 12.35}          & 40.93          & \multicolumn{2}{c}{90.33 / 14.83}           & 61    & \textbf{43.36}  & \multicolumn{2}{c}{\textbf{90.38 / 16.90}}  & 52    & \underline{45.04}     & \multicolumn{2}{c}{90.06 / 16.79}       & 39       \\
      notebook2             & \multicolumn{2}{c}{\textbackslash{}}       & 41.47          & \multicolumn{2}{c}{90.12 / 3.70}            & 55    & \textbf{41.55}  & \multicolumn{2}{c}{\textbf{90.37 / 3.97}}   & 52    & \underline{41.96}     & \multicolumn{2}{c}{90.05 / 2.82}        & 42       \\
      altosight             & \multicolumn{2}{c}{50.22 / 15.30}          & \textbf{46.04} & \multicolumn{2}{c}{\textbf{90.69 / 26.67}}  & 27    & 38.83           & \multicolumn{2}{c}{90.32 / 13.74}           & 51    & 44.00           & \multicolumn{2}{c}{90.54 / 21.89}       & 24       \\
      altosight2            & \multicolumn{2}{c}{\textbackslash{}}       & 28.50          & \multicolumn{2}{c}{\textbf{90.03 / 3.51}}   & 85    & \textbf{29.04}  & \multicolumn{2}{c}{89.80 / 2.81}            & 100   & 28.46           & \multicolumn{2}{c}{\underline{90.19 / 3.65}}  & 49       \\ \cmidrule(lr){1-1} \cmidrule(lr){2-3} \cmidrule(lr){4-7} \cmidrule(lr){8-11} \cmidrule(lr){12-15}
      fodors-zagats\_homo   & \multicolumn{2}{c}{100.00 / 4.73}          & \textbf{60.51} & \multicolumn{2}{c}{\textbf{100.00 / 21.01}} & 1     & \textbf{60.51}  & \multicolumn{2}{c}{\textbf{100.00 / 21.01}} & 1     & 56.41           & \multicolumn{2}{c}{100.00 / 12.81}      & 1        \\
      dblp-acm              & \multicolumn{2}{c}{100.00 / 4.18}          & 91.56          & \multicolumn{2}{c}{98.74 / 83.94}           & 1     & \textbf{91.88}  & \multicolumn{2}{c}{\textbf{99.19 / 84.33}}  & 1     & 87.81           & \multicolumn{2}{c}{99.69 / 75.87}       & 1        \\
      dblp-scholar          & \multicolumn{2}{c}{100.00 / 0.10}          & \textbf{74.27} & \multicolumn{2}{c}{\textbf{92.16 / 31.40}}  & 6     & 73.12           & \multicolumn{2}{c}{91.55 / 31.19}           & 6     & 71.60           & \multicolumn{2}{c}{92.24 / 26.17}       & 5        \\
      abt-buy\_homo         & \multicolumn{2}{c}{89.00 / 14.90}          & \textbf{80.58} & \multicolumn{2}{c}{92.80 / 31.39}           & 3     & 79.80           & \multicolumn{2}{c}{\textbf{93.16 / 31.51}}  & 3     & 80.35           & \multicolumn{2}{c}{\underline{94.26 / 36.38}} & 2        \\
      walmart-amazon\_homo  & \multicolumn{2}{c}{99.77 / 0.27}           & \textbf{56.93} & \multicolumn{2}{c}{\textbf{90.47 / 20.44}}  & 2     & 53.22           & \multicolumn{2}{c}{93.33 / 14.06}           & 3     & 54.92           & \multicolumn{2}{c}{94.37 / 14.26}       & 2        \\ \cmidrule(lr){1-1} \cmidrule(lr){2-3} \cmidrule(lr){4-7} \cmidrule(lr){8-11} \cmidrule(lr){12-15}
      fodors-zagats\_heter  & \multicolumn{2}{c}{100.00 / 3.64}          & \textbf{59.52} & \multicolumn{2}{c}{\textbf{98.21 / 20.64}}  & 1     & 59.35           & \multicolumn{2}{c}{\textbf{98.21 / 20.64}}  & 1     & 56.31           & \multicolumn{2}{c}{100.00 / 12.61}      & 1        \\
      imdb-dbpedia          & \multicolumn{2}{c}{52.91 / 5.46}           & \textbf{37.34} & \multicolumn{2}{c}{\textbf{69.96 / 0.58}}   & 100   & 23.93           & \multicolumn{2}{c}{59.53 / 0.49}            & 100   & 34.35           & \multicolumn{2}{c}{\underline{71.43 / 0.31}}  & 100      \\
      amazon-google         & \multicolumn{2}{c}{94.87 / 3.90}           & 61.60          & \multicolumn{2}{c}{\textbf{91.62 / 17.48}}  & 5     & \textbf{62.31}  & \multicolumn{2}{c}{90.38 / 17.24}           & 5     & \underline{64.05}     & \multicolumn{2}{c}{\underline{92.54 / 20.30}} & 3        \\
      abt-buy\_heter        & \multicolumn{2}{c}{90.05 / 14.80}          & 80.85          & \multicolumn{2}{c}{\textbf{90.06 / 45.70}}  & 2     & \textbf{80.89}  & \multicolumn{2}{c}{93.16 / 31.51}           & 3     & \underline{81.24}     & \multicolumn{2}{c}{94.99 / 36.38}       & 2        \\
      walmart-amazon\_heter & \multicolumn{2}{c}{99.73 / 0.01}           & \textbf{33.85} & \multicolumn{2}{c}{80.68 / 0.36}            & 100   & 30.18           & \multicolumn{2}{c}{\textbf{90.03 / 0.75}}   & 54    & \underline{38.18}     & \multicolumn{2}{c}{\underline{90.03 / 1.11}}  & 20       \\ \cmidrule(lr){1-1} \cmidrule(lr){2-3} \cmidrule(lr){4-7} \cmidrule(lr){8-11} \cmidrule(lr){12-15}
      movies                & \multicolumn{2}{c}{87.17 / 0.27}           & \textbf{14.48} & \multicolumn{2}{c}{\textbf{90.37 / 2.35}}   & 12    & 8.07            & \multicolumn{2}{c}{90.04 / 2.20}            & 13    & 12.00           & \multicolumn{2}{c}{\underline{90.45 / 3.90}}  & 5        \\ \cmidrule(lr){1-1} \cmidrule(lr){2-3} \cmidrule(lr){4-7} \cmidrule(lr){8-11} \cmidrule(lr){12-15}
      Mean                  & \multicolumn{2}{c}{86.31 / 10.42}          & \textbf{51.97} & \multicolumn{2}{c}{89.73 / 22.58}           & 33.24 & 49.60           & \multicolumn{2}{c}{\textbf{89.86 / 20.29}}  & 32.35 & 50.98           & \multicolumn{2}{c}{\underline{91.72 / 19.52}} & 23.53    \\ \bottomrule
    \end{tabular}
  }
\end{table*}
\begin{table}[]
\end{table}


\subsubsection*{\sys vs Unified Sentence-Based Models.}
To explore the necessity of pre-training on structured data, we compare \sys with the sentence embedding language model.
The performance of STransformer, which served as our initialized model for pre-training, is detailed in \autoref{tab:dense}. Our experimental results show that \sys outperforms STransformer by over 10\% mAP on average. STransformer performs relatively poorly on some less challenging but more structured datasets, such as ``fodors-zagats\_homo'', ``abt-buy\_homo'', and ``walmart-amazon\_homo''. This highlights the barriers to applying language models directly to blocking due to the divergences between unstructured natural language and structured records.
Overall, our results emphasize the necessity of training the universal blocker on structured and semi-structured records.

\emphasis{Finding 2. \sys outperforms sentence-based models, supporting the need to align unstructured language models with structured and semi-structured records by pre-training.}

\subsubsection*{Domain-Specific Fine-tuning for \sys.}
To investigate the potential benefits of domain-specific fine-tuning for \sys, we further fine-tune \sys in the one-model-per-dataset paradigm. The mAP changes of the fine-tuned models are also presented in \autoref{tab:dense}. Comparing the experimental results with and without fine-tuning, we observe that \sys derives limited benefit from fine-tuning except for ``imdb-dbpedia''. This finding suggests that the domain-independent pre-training of \sys has resulted in a universal blocking capability that is sufficient to address domain-specific blocking challenges without the need for further fine-tuning. Consequently, this finding reaffirms the effectiveness of \sys in solving blocking problems across multiple domains.

\emphasis{Finding 3. \sys benefits little from domain-specific self-supervised fine-tuning, indicating that \sys can learn sufficient knowledge from domain-independent pre-training.}

\subsection{Comparison with Sparse Blocking Methods}

\subsubsection*{Effectiveness Evaluation.}
Initially, we compare the effectiveness of \sys and SOTA sparse blocking methods.
Our experimental results, shown in \autoref{tab:dense-vs-sparse}, reveal that Sparkly surpasses \sys by about 2\% mAP on average.
However, it's worth noting that \sys is comparable to Sparkly on simple homogeneous datasets, and blocking better on some more difficult datasets, such as ``walmart-amazon\_heter'' and ``notebook''.
In addition, the ensemble of \sys and Sparkly can improve PC by up to 5\% on dataset ``cora''.
Lastly, self-supervised dense retrieval models have outperformed BM25 as more high-quality data become available and model parameters expand. This suggests that the performance of \sys will continue to improve over time, showing great promise for universal self-supervised dense blocking.

\emphasis{Finding 4. \sys is comparable to SOTA sparse blocking methods and they complement each other to a certain extent.}

\begin{figure}
  \setlength{\abovecaptionskip}{0pt}
  \centering
  \includegraphics[width=0.9\columnwidth]{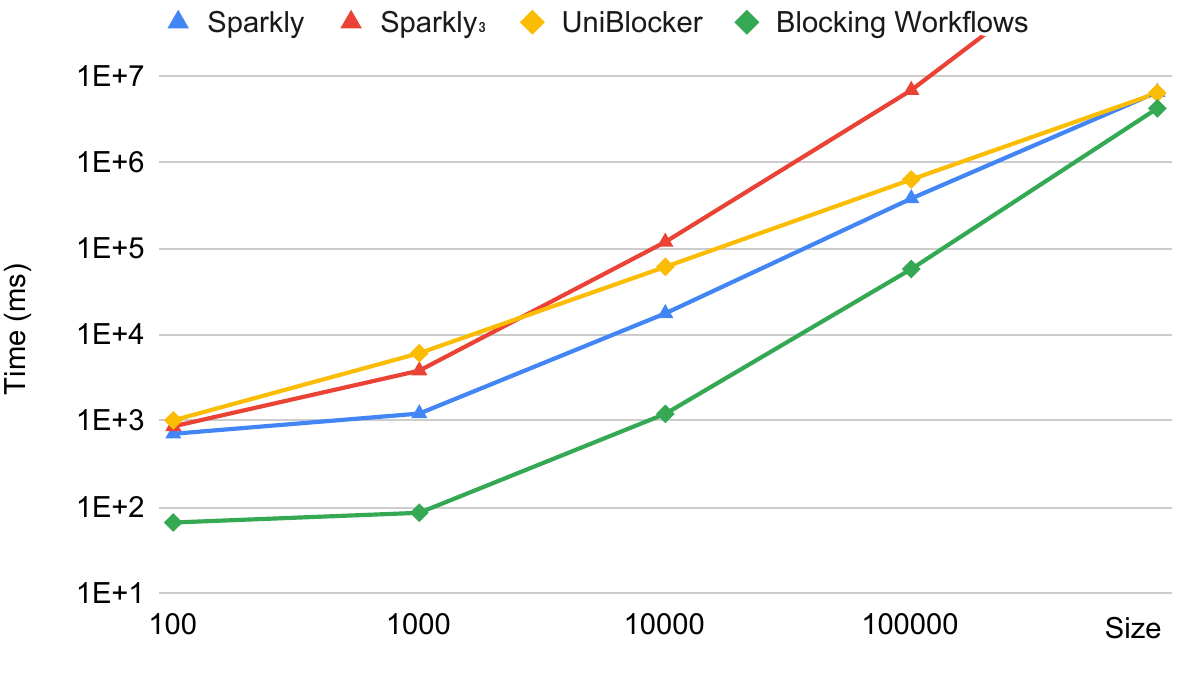}
  \caption{Average blocking times for varying approaches on scalability datasets with different record sizes.}
  \label{fig:total-time}
\end{figure}

\subsubsection*{Efficiency Analysis.}
Furthermore, we perform efficiency analysis on scalability datasets from our proposed benchmark.
To this end, we downsample the entity collections from $10^2$ to $10^6$ and compare the blocking time of different approaches. As shown in \autoref{fig:total-time}, sparse blocking methods are efficient for small numbers of records, but the difference decreases as the number of records increases.
We can see that \sys has the slowest time growth curve because it generates fixed-length dense representations for records. However, for Sparkly with a 3-gram tokenizer, the number of tokens and the number of record pairs with the same tokens grow non-linearly, posing challenges for index storage and nearest neighbor search. Therefore, q-gram technology improves the effectiveness but sacrifices some scalability at larger scales.

\emphasis{Finding 5. Sparse blocking methods are more efficient than \sys for small numbers of records, but the difference decreases as the number of records increases.}

\subsection{Ablation Study}
We now perform ablation studies on \sys to validate the importance of specific components, as depicted in \autoref{tab:ablation}.

\begin{table}[]
  \centering
  \caption{Average performance in ablation studies. \sys ``w/o OD'' means pre-training with sufficient out-domain data instead of open-domain GitTables. ``RE'', ``LP'', ``FP'', and ``DD'' stand for record embedding, literal paraphrasing, feature paraphrasing, and duplicate detection, respectively.}
  \label{tab:ablation}
  \resizebox{0.7\linewidth}{!}{
    \begin{tabular}{@{}lcccc@{}}
      \toprule
      Methods                   & mAP   & PC    & PQ    & k     \\ \midrule
      \sys w/o OD         & 46.48 & 89.37 & 18.00 & 35.76 \\ \cmidrule{1-5}
      \sys w/o RE         & 48.31 & 89.29 & 19.72 & 33.82 \\ \cmidrule{1-5}
      \sys w/o LP         & 49.09 & 89.40 & 20.23 & 34.71 \\
      \sys w/o FP         & 48.64 & 89.49 & 19.75 & 35.94 \\
      \sys w/o DD         & 48.31 & 89.06 & 19.65 & 36.12 \\ \cmidrule{1-5}
      \sys                & 49.60 & 89.86 & 20.29 & 32.35 \\ \bottomrule
    \end{tabular}%
  }
\end{table}

\emph{Effectiveness of Pre-training Data.}
To explore the effect of pre-training data diversity on \sys, we replace the training data extracted from open-domain GitTables with out-domain tables sourced from scalability datasets in our benchmark. This helps to establish whether the pre-training data serves solely to bridge the gap between natural language and structured records, or if it also contributes to the universality of \sys. Experimental results in \autoref{tab:ablation} show that \sys achieves a significant performance gain over the model trained on out-domain data, suggesting that open-domain data pre-training is crucial for its universality.

\emph{Effectiveness of Record Embedding.}
After replacing the proposed record embedding (RE) with the previous serialization way, a performance gap of 1.29\% on mAP is observed, as presented in \autoref{tab:ablation}.
We believe that the limitations of the prior serialization way have been concealed, as it was mainly used for entity matching where cross-record interactions occur within the same model.
For blocking, however, only one record is passed to the model at a time.
The inclusion of extra tokens and attribute names can impact the final representation of records, particularly when attribute values are short or highly missing.
The proposed Record Embedding, therefore, can help Transformer-based architectures to better represent a single record and be more suitable for blocking.

\emph{Effectiveness of Data Paraphrasing.}
We then experiment with the effect of different data paraphrasing approaches by removing them in turn. As shown in \autoref{tab:ablation} again, we find that both feature paraphrasing (FP) and lexical paraphrasing (LP) contribute to the performance of \sys, among which LP is more pronounced because this type of paraphrasing is more refined, diverse, and stable, without the chance of introducing false positives. Moreover, due to the complexity of the real data, the application of duplicate detection (DD) achieves a 1.29\% mAP improvement.

\emphasis{Finding 7. Pre-training on open-domain data and the techniques applied in \sys are effective and necessary.}

\section{Related Work}
\label{sec:related work}

\subsection{Blocking}

Blocking, a crucial step in addressing the inherently quadratic complexity of entity resolution, has been approached in numerous ways~\cite{papadakis-21-block-filter-techn-entit-resol}.
Blocking Workflows~\cite{papadakis-20-three-entit-resol-jedai}, which originated in the late 1960s~\cite{fellegi-69-theor-recor-linkag}, involve block building~\cite{fellegi-69-theor-recor-linkag,papadakis2011eliminating,zhang-20-autob}, block cleaning~\cite{papadakis2013blocking,papadakis2015schema}, and comparison cleaning~\cite{papadakis2013blocking,papadakis2014meta} to group records into blocks using signatures derived from records. However, challenges arise in precisely controlling candidate pair numbers, tuning pipeline~\cite{li-18-match,galhotra-21-beer}, and automating the tuning~\cite{papadakis-22-how-entit-resol}.
Sparse join~\cite{gravano2001approximate,paulsen-23-spark} operates by identifying syntactically similar pairs in collections using either threshold-based~\cite{chaudhuri2006primitive} or cardinality-based~\cite{bohm2002high} conditions within a filter-verification framework. While effective, they sometimes struggle with loose conditions, necessitating approximation techniques~\cite{gionis1999similarity,zhai2011atlas,malkov2020efficient} for scalability.
Dense blocking represents records as dense vectors and performs similarity searches to find potential pairs~\cite{ebraheem-18-distr}. Despite improving record representation through the use of pre-trained language models~\cite{li-20-deep-entit-match-pre-train-languag-model} or via self-supervised~\cite{thirumuruganathan-21-deep,wang-22-sudow} and active learning~\cite{jain-21-deep}, performance is constrained by the divergences between natural language and structured records and the absence of record-specific pre-trained models.

\subsection{Table Pre-training}

Recent developments in table pre-training, motivated by the success of pre-training paradigms, have significantly impacted various downstream tasks~\cite{dong-22-table-pre}. Early stages used pre-training word embeddings on relational or web datasets~\cite{cappuzzo-20-creat-embed-heter-relat-datas,gunther2021pre}, which later evolved to Transformer-based language models.
In the area of natural language processing, the focus is on tasks integrating tabluar and textual data, such as table question answering~\cite{yin2020tabert,herzig2020tapas}, fact verification~\cite{eisenschlos2021mate,liu2021tapex}, and table summarization~\cite{xing2021structure}, which typically take whole tables as input.
In terms of data integration and preparation, the emphasis is on table-centric tasks such as table interpretation, cleaning, and argumentation, with most studies~\cite{deng-20-turl,iida2021tabbie,wang2021tuta} pre-training models on web tables and supervised fine-tuning on specific tasks. The advent of data lakes has sparked interest in table union search, with contrastive pre-training frameworks~\cite{fan2022semantics,dong2022deepjoin,cong2023pylon} taking table columns as input.
In the field of machine learning, there have been efforts~\cite{yoon2020vime,wang2022transtab} to compare the performance of pre-trained deep models and traditional methods in classification and regression tasks on tabular tables comprising categorical and continuous features.
Notably, none of these works pre-trained models for entity record representations and can be directly applied to blocking.


\section{Conclusion}
\label{sec:conclusion}

The main focus of this paper is to explore the potential of developing a universal dense blocker that can be applied across diverse scenarios and domains through large-scale self-supervised learning. To achieve this, we propose a unified contrastive pre-training framework that exploits the power of cross-domain pre-training for developing the universal dense blocker, \sys. The pre-training is performed via self-supervised contrastive learning on large-scale relational tables. We also introduce a new record embedding technique that allows Transformer-based architectures to better model individual records. In addition, we propose three data paraphrasing techniques to generate positives for training. To assess the effectiveness and universality of \sys, we create a new benchmark for universal blocking evaluation that covers a wide range of blocking tasks from multiple domains and scenarios. Experimental results demonstrate the effectiveness and universality of \sys, holding great promise for universal dense blocking.



\bibliographystyle{ACM-Reference-Format}
\bibliography{ref/core,ref/main,ref/other}

\clearpage

\appendix
\begin{algorithm}[!t]
  \caption{Heuristic Rule for Column Type Detection.}
  \label{alg:column type detection}
  \KwIn{A table column $col$ with its name $col.name$ and list of values $col.values$.}
  \KwOut{One of the following predefined column types:
  $\left\{ \textsf{identifier}, \textsf{numeric}, \textsf{url}, \textsf{date}, \textsf{category}, \textsf{entity}, \textsf{text} \right\}$}
  \If{number of unique values in $col.values$ is 100\% of its length}{
    \Return $\textsf{identifier}$
  }
  \If{$col.values$ are numeric or boolen}{
    \Return $\textsf{numberic}$
  }
  \If{$col.values$ are url}{
    \Return $\textsf{url}$
  }
  \If{$col.values$ are datetime}{
    \Return $\textsf{date}$
  }
  $value\_lens$ $\leftarrow$ number of token of $col.values$ \\
  \eIf{variance of $value\_lens$ is 0}{
    \Return $\textsf{category}$
  }{
    \eIf{average of $value\_len$ less than 10}{
        \Return $\textsf{entity}$
    }{
        \Return $\textsf{text}$
    }
  }
\end{algorithm}

\section{Details of Pre-training Data Conditioning}

This section further introduces the details of pre-training data conditioning:

\textbf{Column Type Detection.} The column type detection algorithm is shown in \autoref{alg:column type detection}, originally used for tabular corpus analysis. Specifically, we define seven column types, namely \texttt{identifier}, \texttt{numeric}, \texttt{URL}, \texttt{date}, \texttt{category}, \texttt{entity}, and \texttt{text}. To begin, we determine whether a given column is an \texttt{identifier} by checking whether the column value selectivity is 100\%. Following this, we assess whether the column values are \texttt{numeric} (either boolean or float). For columns representing \texttt{URL} or \texttt{date} information, we utilize regular expressions and the Python package dateutil\footnote{\url{https://github.com/dateutil/dateutil}} to match patterns. We segregate the remaining columns into either \texttt{category} or \texttt{entity} and \texttt{text} types, depending on whether the length variance of column values is equal to zero. We differentiate \texttt{entity} from \texttt{text} by examining if the number of tokens is less than 10.

\textbf{Table filtering.} After the column type detection, we apply additional filtering to exclude tables that do not contain suitable entity records. We randomly select 100 tables for manual labeling and train a decision tree using features like the number of each column type and the type of the first column. To make the filtering algorithm more robust, we refine the decision tree manually to create the final filtering algorithm. Specifically, we filter out tables containing statistics where more than half of the columns are comprised of numeric, URL, or date information. We also exclude tables where the first column is a date, which is typically associated with log data. Finally, we filter tables that are not written in English.

\section{Benchmark Discussion}

Our proposed benchmark value can be further validated by experimental results. On the one hand, \autoref{tab:dense-vs-sparse} shows that existing blocking methods can effectively address the homogeneous and simple datasets that previous studies mainly focused on in \autoref{tab:benchmark}. There is still room for improvement on some of our newly introduced datasets, providing new challenges and opportunities for future research.
On the other hand, our proposed benchmark is the first to evaluate the general performance of different methods in different scenarios using the same parameters on all datasets, laying the foundation for future automatic, data-driven sparse blocking tuning approaches or good attribute identification for blocking.
Finally, by comparing the performance differences of \sys on artificially and naturally heterogeneous datasets with their corresponding homogeneous datasets, we can see that the mainstream approach of artificially constructing dirty data does not pose much of a challenge to deep models. How to better perform entity resolution directly on real highly heterogeneous data (such as ``walmart-amazon\_heter'') is also a future research direction.

\section{Nearest Neighbor Blocker}

\begin{algorithm}[!t]
  \DontPrintSemicolon
  \caption{Nearest Neighbor Blocker}
  \label{alg:nnblocker}
  \KwIn{list of entity record collections $collections$, record vectorizer $vectorizer$, nearest neighbor indexer $indexer$, and nearest neighbor number $k$}
  \KwOut{list of candidate pair sets arranged in nearest neighbor ranking $candidates$}
  \If{number of $collections > 2$}{
    merge $collections$ into one
  }
  $source \leftarrow \operatorname{vectorize}(collections[0], vectorizer)$ \;
  $target \leftarrow \operatorname{vectorize}(collections[-1], vectorizer)$ \;
  $index \leftarrow \operatorname{buildIndex}(target, indexer)$ \;
  $indices \leftarrow []$ \tcp*[r]{empty list}
  \For{$batch\_queries$ $\mathrm{in}$ $\operatorname{chunk}(source)$}{
    $batch\_indices \leftarrow \operatorname{batchSearch}(batch\_queries, index, k)$ \;
    extend $batch\_indices$ to the end of $indices$ \;
  }
  \tcp{convert $indices$ to list of candidate pair sets arranged in nearest neighbor ranking $candidates$}
  $candidates \leftarrow []$ \;
  \For{$i$ $\mathrm{in}$ $\operatorname{range}(k)$}{
    $candiate\_pairs \leftarrow \{\}$ \tcp*[r]{empty set}
    \For{$j$ $\mathrm{in}$ $\operatorname{range}(\operatorname{len}(source))$}{
      $candidate\_pair \leftarrow (j, indices[j][i])$ \;
      \If{$candidates\_pair$ did not appear before}{
        insert $candidate\_pair$ into $candidate\_pairs$ \;
      }
    }
    append $candidate\_pairs$ to the end of $candidates$ \;
  }
  \Return $candidates$
\end{algorithm}

This section presents how to use \sys for blocking any number of entity record collections. It is worth mentioning that our general implementation of the nearest neighbor blocker can apply different sparse or dense vectorizers and indexers for a thorough and fair comparison of different methods. As shown in \autoref{alg:nnblocker}, the nearest neighbor blocker operates on a list of entity collections, using a record vectorizer and a nearest neighbor indexer to identify $k$ nearest neighbors for each record. In cases with more than two collections, they are unified into a single one. The first and last collections are vectorized as the $source$ and $target$ respectively, with the latter being indexed. The algorithm iteratively searches for $k$ nearest neighbors of source vectors and stores these indices. Subsequently, candidate pairs are formulated, each constituted of a source index and its corresponding nearest neighbor index from the target. Each candidate pair is prioritized and deduplicated according to its nearest neighbor ranking, to facilitate progressive entity matching or evaluation. This sequence of candidate pair sets forms the final output of the algorithm.

Besides offering a more precise assessment for blocking, our suggested benchmark value can be further corroborated by experimental results. On one hand, \autoref{tab:dense-vs-sparse} demonstrates that current sparse and dense blocking methods effectively tackle homogeneous and simple datasets primarily studied in \autoref{tab:benchmark}. However, there is scope for improvement on some newly introduced datasets, which present new challenges and opportunities for future research. On the other hand, our suggested benchmark is the first to evaluate general performance across various scenarios using the same parameters for all datasets, setting the groundwork for future data-driven sparse blocking tuning approaches~\cite{papadakis-22-how-entit-resol} or more versatile dense blocking model evaluations. Lastly, by comparing \sys's performance differences on artificially and naturally heterogeneous datasets with their homogeneous counterparts, we observe that artificial dirty data construction does not significantly challenge deep models. A future research direction includes improving entity resolution on highly heterogeneous real data (such as "walmart-amazon\_heter").

\begin{figure}
  \centering
  \caption{Average number of tokens using different tokenizers with increasing entity record size on two scalability datasets.}
  \label{fig:token number}
  \includegraphics[width=\columnwidth]{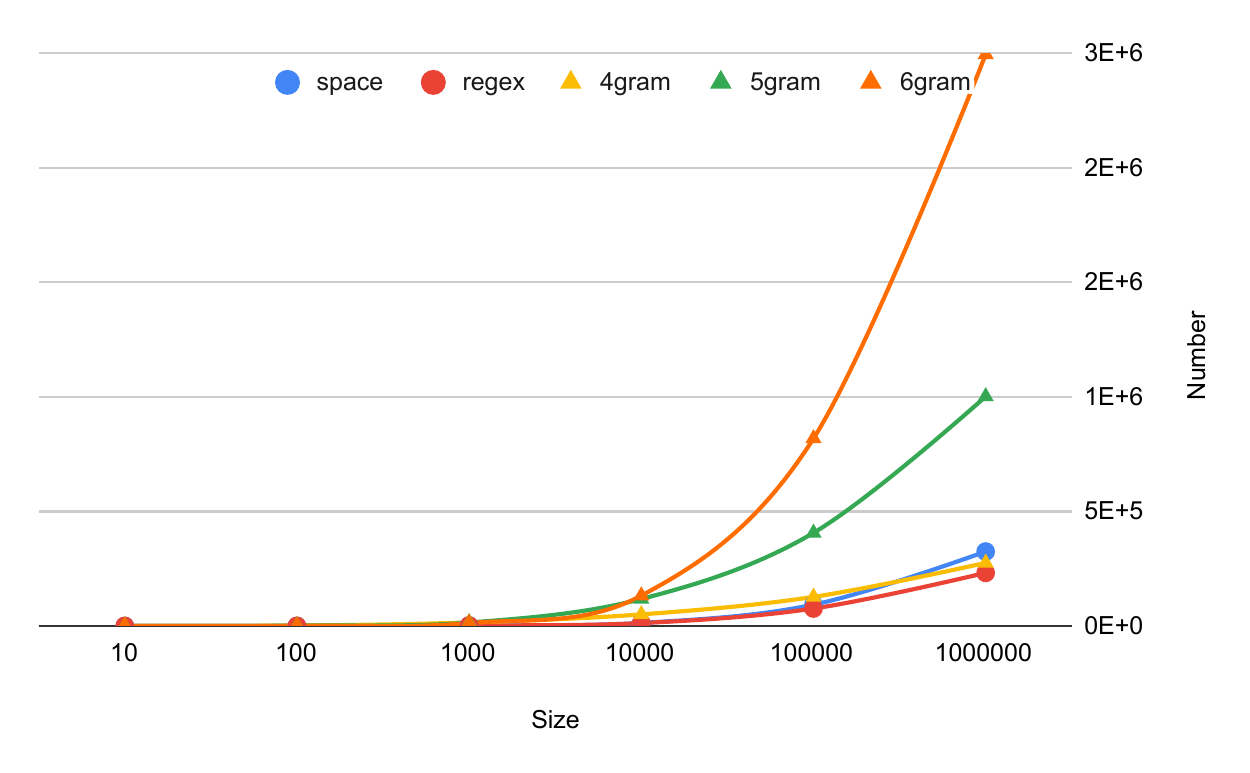}
\end{figure}

\begin{figure}
  \centering
  \caption{Average number of overlapping records (i.e. records sharing at least one common token) using different tokenizers with increasing entity record size on two scalability datasets.}
  \label{fig:overlapping records}
  \includegraphics[width=\columnwidth]{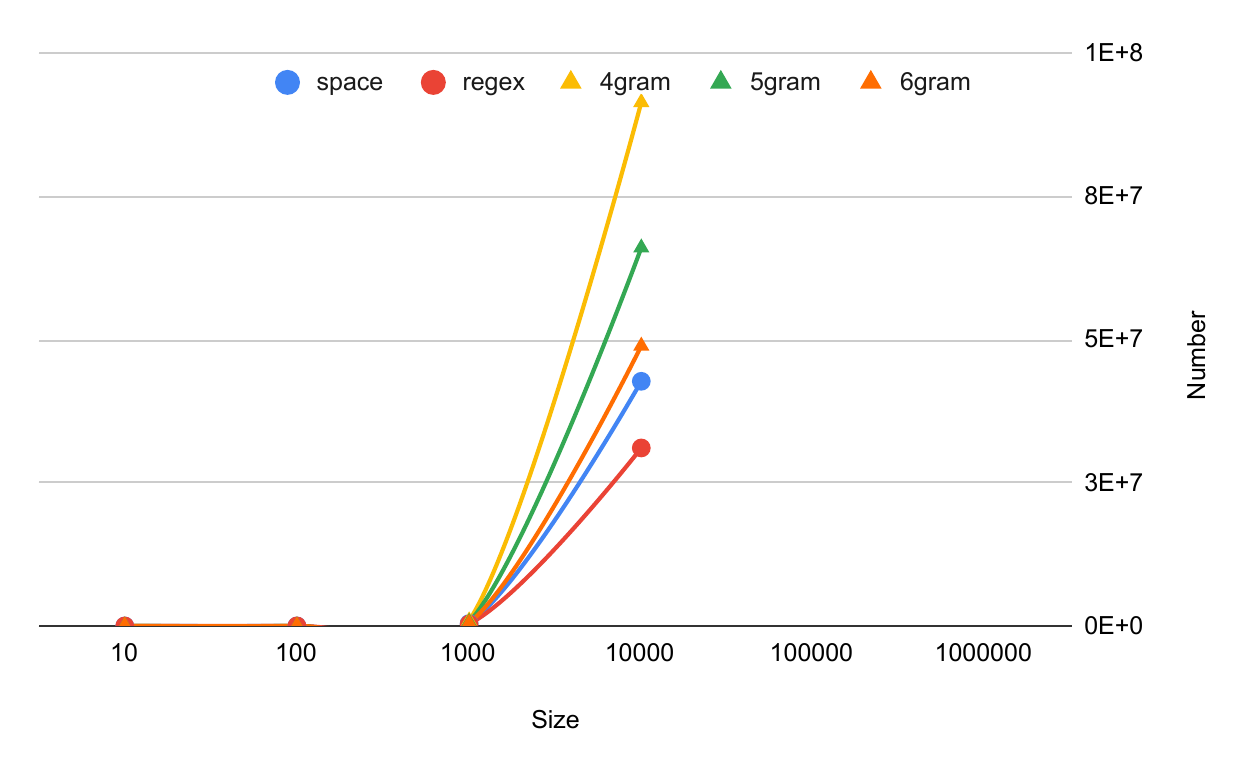}
\end{figure}

\section{Impact of Tokenizers on Tokens and Overlapping Records.}

Although q-gram is an effective fuzzing matching technique, it sacrifices blocking scalability. In this section, we present a concise overview of the impact of varying tokenizers on the number of tokens and overlapping records (i.e. records sharing at least one common token). As shown in \autoref{fig:token number}, the q-gram tokenizers produce far more tokens than the regular tokenizers do. The ideal upper bound on the number of tokens produced by the q-gram tokenizer is the $m^q$, where m is the number of characters. With the increase of q, the growth rate of token number will accelerate rapidly. Exponentially growing tokens pose a serious challenge to the size and speed of building index, which is what the filter-verification framework needs. Futhermore, though a smaller q produces fewer tokens, as shown in \autoref{fig:overlapping records}, this produces more overlapping records. Sub-quadratic record pairs with shared tokens may also pose a challenge for finding similar records. As a result, dense blocking with fixed dimension vectors not only models semantic information but is also more scalable than q-gram sparse blocking methods.

\begin{table*}
  \centering
  \caption{Overall performance of sparse, dense kNN join with approximate nearest neighbor search.}
  \label{tab:dense-vs-sparse-ann}
  \resizebox{0.7\textwidth}{!}{%
    \begin{tabular}{@{}r|cccccccccc@{}}
      \toprule
      & \multicolumn{5}{c}{Sparse kNN}        & \multicolumn{5}{c}{Dense kNN}         \\ \cmidrule(lr){2-6} \cmidrule(lr){7-11}
      Datasets              & AP    & PC    & PQ    & F1    & k     & AP    & PC    & PQ    & F1    & k     \\ \midrule
      census                & 69.89 & 90.34 & 15.28 & 26.14 & 6     & 73.17 & 90.88 & 23.06 & 36.78 & 4     \\
      cora                  & 66.57 & 90.25 & 9.16  & 16.63 & 10    & 73.21 & 90.06 & 22.85 & 36.45 & 4     \\
      notebook              & 39.08 & 90.15 & 14.96 & 25.67 & 43    & 36.59 & 90.13 & 11.19 & 19.90 & 59    \\
      notebook2             & 24.45 & 82.92 & 2.48  & 4.81  & 100   & 26.22 & 83.36 & 2.55  & 4.95  & 100   \\
      altosight             & 49.12 & 90.08 & 1.95  & 3.82  & 44    & 48.03 & 90.15 & 3.74  & 7.18  & 23    \\
      altosight2            & 11.89 & 93.60 & 9.24  & 16.82 & 7     & 20.42 & 95.06 & 15.47 & 26.61 & 5     \\ \midrule
      fodors-zagats\_homo   & 47.37 & 71.46 & 34.73 & 46.74 & 100   & 48.01 & 72.04 & 34.44 & 46.60 & 100   \\
      dblp-acm              & 89.04 & 96.63 & 82.15 & 88.80 & 1     & 90.94 & 98.29 & 83.56 & 90.33 & 1     \\
      dblp-scholar          & 67.23 & 90.03 & 26.29 & 40.69 & 7     & 68.44 & 90.99 & 23.25 & 37.03 & 8     \\
      abt-buy\_homo         & 55.74 & 92.86 & 19.51 & 32.25 & 1     & 58.79 & 97.32 & 20.45 & 33.80 & 1     \\
      walmart-amazon\_homo  & 57.98 & 96.43 & 20.26 & 33.49 & 1     & 59.95 & 99.11 & 20.83 & 34.42 & 1     \\ \midrule
      fodors-zagats\_heter  & 12.83 & 30.69 & 0.28  & 0.56  & 100   & 6.86  & 26.28 & 0.22  & 0.43  & 100   \\
      imdb-dbpedia          & 3.11  & 75.28 & 0.20  & 0.40  & 100   & 4.80  & 85.37 & 0.25  & 0.50  & 100   \\
      amazon-google         & 31.65 & 90.06 & 11.15 & 19.84 & 70    & 34.85 & 90.06 & 8.50  & 15.53 & 98    \\
      abt-buy\_heter        & 40.44 & 90.12 & 2.61  & 5.07  & 72    & 38.28 & 90.16 & 3.10  & 5.99  & 65    \\
      walmart-amazon\_heter & 29.76 & 85.18 & 0.38  & 0.77  & 100   & 25.99 & 83.19 & 0.38  & 0.75  & 100   \\ \midrule
      movies                & 48.76 & 90.47 & 4.09  & 7.82  & 10    & 47.81 & 90.12 & 6.79  & 12.63 & 6     \\
      Average               & 43.82 & 85.09 & 14.98 & 21.78 & 45.41 & 44.84 & 86.03 & 16.51 & 24.11 & 45.59 \\ \bottomrule
    \end{tabular}
  }
\end{table*}

\section{Impact of ANN Search on Effectiveness}

\autoref{tab:dense-vs-sparse-ann} shows the overall performances of sparse and dense blocking methods with the same approximate nearest neighbor algorithm (ANN) and configuration. In this simple setting, we can find that the average mAP of the sparse blocking method decreases by 3.09 compared to the precision search, but the dense blocking method only decreases by 1. Therefore, sparse join may receive more influence from ANN search than dense join.


\end{document}